\long\def\comment#1{}
\let\OldTheta=\Theta 
\documentclass{article}
\usepackage{amssymb}
\newdimen{\pxh}
\newdimen{\pxv}
\newcount{\BinPicN}
\def\binpic#1{{%
\def\h{1}%
\def\L##1{\hbox{{\R0}##1}\nointerlineskip}%
\def\V##1{\vskip##1\pxh\relax}%
\def\R##1{\vrule width##1\pxh height\h\pxv depth0pt}%
\def\S##1{\hskip##1\pxh\relax}%
\def\H##1##2{{\def\h{##1}{##2}}}%
\def\N##1##2{{\BinPicN=\number##1\loop\ifnum\BinPicN>0{##2}\advance\BinPicN by-1\repeat}}%
\def\0{\S4}\def\1{\S3\R1}\def\2{\S2\R1\S1}\def\3{\S2\R2}%
\def\4{\S1\R1\S2}\def\5{\S1\R1\S1\R1}\def\6{\S1\R2\S1}\def\7{\S1\R3}%
\def\8{\R1\S3}\def\9{\R1\S2\R1}\def\A{\R1\S1\R1\S1}\def\B{\R1\S1\R2}%
\def\C{\R2\S2}\def\D{\R2\S1\R1}\def\E{\R3\S1}\def\F{\R4}%
\hbox{\vbox{#1}}%
}}%
\def\picscale#1#2#3{{%
\ifx#1?\relax\else\pxh=#1\fi%
\ifx#2?\relax\else\ifx#2*\pxv=\pxh\else\pxv=#2\fi\fi%
#3%
}}%
\def\qisSPb{%
\picscale{0.24bp}{0.24bp}{\binpic{%
\L{\S{56}\8\0}%
\L{\S{48}\0\1\8\0}%
\L{\S{48}\3\F\8\0}%
\L{\S{48}\2\0\8\0}%
\L{\S{48}\C\7\8\0}%
\L{\S{48}\8\C\8\0\S{16}\0\1}%
\L{\S{48}\F\8\8\0\S{16}\0\1}%
\L{\S{48}\8\0\8\0\S{16}\0\1}%
\L{\S{56}\8\0\S{16}\0\1}%
\L{\S{16}\0\2\S{32}\8\0\S{16}\0\1}%
\L{\S{16}\0\2\S{32}\8\0\S{16}\0\1}%
\L{\S{16}\0\2\S{32}\8\0\S{16}\0\1}%
\L{\S{16}\0\2\S{32}\8\0\S{16}\0\1}%
\L{\S{16}\0\2\S{32}\8\0\S{16}\0\1}%
\L{\S{16}\0\2}%
\L{\S8\0\1\C\2\0\7\S{16}\3\F\C\3\S{16}\3\F\F\C}%
\L{\S8\0\7\8\2\0\F\C\0\S8\7\F\F\E\S8\0\1\C\1\R8\C\0}%
\L{\S8\0\E\S8\1\F\E\0\S8\C\F\F\C\S8\0\1\S8\3\F\8\8}%
\L{\S8\0\C\S8\6\9\F\0\0\1\0\3\F\8\0\8\0\2\S8\0\F\8\4}%
\L{\S8\1\C\S8\8\8\F\0\S{24}\1\0\0\6\S8\0\7\0\4}%
\L{\S8\3\8\0\F\0\8\F\8\S{24}\3\0\0\6\S8\0\2\0\2}%
\L{\S8\3\8\3\C\0\8\7\C\S{24}\6\0\0\E\S8\0\4\0\1}%
\L{\S8\7\8\3\C\0\8\7\C\S{24}\6\0\0\7\S8\0\8\0\1}%
\L{\S8\7\0\3\C\0\8\3\E\S{16}\2\0\E\0\0\7\C\0\1\0\S8\8\0}%
\L{\S8\7\0\3\C\0\8\1\E\S{16}\6\0\E\0\0\3\R{16}\F\8\4\0}%
\L{\S8\E\0\3\C\0\8\1\E\S{16}\6\1\E\0\0\1\R{16}\F\E\2\0}%
\L{\S8\E\0\3\C\0\8\0\F\S{16}\E\1\E\0\0\1\R{24}\2\0}%
\L{\0\1\E\0\3\C\0\E\3\F\S{16}\E\1\E\0\S8\7\F\R{16}\9\0}%
\L{\0\1\E\0\3\C\0\9\C\F\S8\0\1\E\1\E\0\S8\0\F\R{16}\8\8}%
\L{\0\1\E\0\3\C\0\8\0\F\S8\0\F\E\1\E\0\2\0\S8\8\0\1\F\C\8}%
\L{\0\1\E\0\3\C\0\8\0\F\S8\1\F\E\1\E\0\3\0\S8\8\0\0\7\C\4}%
\L{\0\1\E\0\3\C\0\8\0\F\S8\3\F\E\1\E\0\3\C\0\1\S8\0\3\C\2}%
\L{\0\1\E\0\3\C\0\8\0\F\S8\3\F\E\1\E\0\1\F\R8\F\E\0\3\C\2}%
\L{\0\1\E\0\3\C\0\8\0\F\S8\3\1\E\1\E\0\1\F\R{16}\0\3\8\1}%
\L{\0\1\E\0\3\C\0\9\C\F\S8\2\1\E\1\E\0\0\F\R{16}\0\3\8\0\8\0}%
\L{\0\1\E\0\3\C\0\E\3\F\S8\0\1\E\1\E\0\0\7\R{16}\0\3\S8\8\0}%
\L{\S8\E\0\3\C\0\8\0\F\S8\0\1\E\1\E\0\0\1\R{16}\0\6\S8\4\0}%
\L{\S8\E\0\3\C\0\8\0\E\S8\0\1\E\1\E\0\S8\1\0\0\F\0\4\S8\2\0}%
\L{\S8\7\0\3\C\0\8\1\E\S8\0\1\C\1\E\0\S8\2\0\0\2\0\C\S8\2\0}%
\L{\S8\7\0\3\C\0\8\1\E\S8\0\1\C\1\E\0\0\8\4\0\0\2\0\8\S8\1\0}%
\L{\0\1\7\8\3\8\0\8\3\C\S8\0\1\8\1\E\0\0\D\8\0\S8\1\0\S8\0\8}%
\L{\0\1\3\8\3\8\0\8\3\C\S8\0\1\0\1\E\0\0\E\S{16}\1\0\S8\0\4}%
\L{\0\2\3\8\2\0\0\8\3\8\S{16}\0\1\E\0\S{24}\2\0\S8\0\4}%
\L{\0\2\1\C\S8\0\8\7\0\S{16}\0\1\C\0\S{24}\2\0\S8\0\2}%
\L{\0\4\0\C\S8\0\8\6\0\0\1\F\8\0\1\C\0\S8\1\F\F\8\4\0\S8\0\1}%
\L{\0\4\0\6\S8\0\8\C\0\0\7\F\E\0\3\8\0\0\1\R{16}\8\0\S8\0\1}%
\L{\0\4\0\3\8\0\0\9\8\0\0\F\R8\8\2\S8\0\3\R{16}\8\8\S{16}\8\0}%
\L{\0\8\S8\C\7\0\E\S8\0\F\R8\F\C\0\8\0\F\R{16}\1\8\S{16}\4\0}%
\L{\0\8\S8\3\4\D\8\S8\1\8\1\F\F\8\0\C\1\F\S8\1\F\1\1\R8\F\8\4\0}%
\L{\1\0\S8\0\F\E\0\0\6\2\0\0\7\E\0\3\E\3\0\S8\0\6\2\0\0\3\8\7\F\8}%
\L{\1\0\0\4\S8\3\0\0\E\0\7\E\0\S8\6\3\S8\7\F\E\0\6\0\0\F\R8\F\0}%
\L{\3\F\E\F\S8\3\8\0\E\0\F\F\8\S8\C\1\S8\E\0\3\0\C\E\1\C}%
\L{\S8\8\7\C\0\1\C\0\F\7\C\7\F\F\B\8\1\R8\8\0\1\C\8\E\1\0}%
\L{\S8\8\3\F\E\1\E\0\7\E\0\0\3\7\E\S8\7\C\S8\0\3\8\0\3\0}%
\L{\S8\6\0\0\F\0\E\0\3\C\0\S8\A\0\S{24}\0\2\S8\4\0}%
\L{\S8\3\0\0\3\8\F\S8\E\1\C\0\S{24}\0\7\3\8\S8\8\0}%
\L{\S8\1\0\S8\C\7\8\0\6\1\2\0\S{16}\1\0\0\7\3\8\0\1}%
\L{\S8\0\C\S8\6\7\E\0\3\1\6\1\2\9\5\0\3\8\S{16}\0\2}%
\L{\S8\0\7\S8\1\3\F\E\3\1\0\2\B\5\8\2\9\0\S{16}\0\C}%
\L{\S{16}\C\0\0\1\F\C\1\1\2\B\2\5\1\3\5\0\S{16}\1\0}%
\L{\S{16}\7\0\S8\7\8\3\1\3\2\3\5\1\2\5\4\S{16}\2\0}%
\L{\S{16}\1\8\S{16}\2\1\2\1\2\8\1\A\4\8\S{16}\4\0}%
\L{\S{16}\0\8\S{16}\4\0\2\0\2\0\0\8\S8\7\0\0\1\8\0}%
\L{\S{16}\0\C\S{56}\8\8\0\3}%
\L{\S8\0\F\E\6\S{48}\0\3\0\7}%
\L{\S8\3\C\1\3\S8\1\8\S{32}\3\C\0\7\R8}%
\L{\S8\1\3\1\D\F\8\0\4\S{32}\1\4\7\8\2\7\E\0}%
\L{\S8\2\0\8\3\F\4\1\F\R8\E\0\S{16}\6\2\4\8\2\0\D\0}%
\L{\S8\2\0\C\2\1\4\2\1\8\0\1\F\F\C\S8\8\3\4\1\A\0\7\F\F\E}%
\L{\S{16}\5\C\0\C\C\0\8\0\S8\0\3\R8\0\1\0\1\C\3\C\F\F\0}%
\L{\S{16}\F\0\0\F\S8\E\0\S8\7\F\R8\E\1\0\1\2\4\0\2}%
\L{\S{16}\9\0\S{16}\1\0\0\1\8\0\S8\3\0\E\6\3\0\0\1\8\0}%
\L{\S{16}\E\0\S{16}\0\F\F\E\S{16}\1\0\1\8}%
}}}%
\def\draftsize{\textheight=230mm\topmargin=-7mm\oddsidemargin=-10mm%
\marginparwidth=40mm}
\def\todo#1{\typeout{"#1" ... maybe later}}
\def\ox{{\otimes}}
\newcommand{\hc}{{*}}

\newcommand{\ket}[1]{| #1 \rangle}
\newcommand{\bra}[1]{\langle #1 |}

\newcommand{\vxv}[1]{\ket{#1}\bra{#1}}
\newcommand{\brkt}[2]{\langle #1 \mid #2 \rangle}
\newcommand\Tr{\mathop\mathrm{Tr}}

\newcommand\Hil{\mathcal H}
\newcommand\C{\mathbb C}
\newcommand\R{\mathbb R}
\newcommand\V{\mathbb V}

\newcommand\mbf[1]{\mbox{\boldmath$#1$}}

\newcommand{\BIG}[3]{{\!\left#1{#2}^{\vphantom{*}}\right#3\!}}

\newcommand{\Abs}[1]{\left|{#1}\right|}
\renewcommand\P{\mathsf\Pi}

\newcommand{\ve}[1]{\mathbf{#1}}

\newcommand\tfrac[2]{{\textstyle\frac#1#2}}
\newcommand\half{{\textstyle\frac12}}
\newcommand\To{\mathrel{{{-}\!{-}\!{\longrightarrow}}}}

\newcommand\Sv{\mathsf S}
\newcommand{\li}[1]{\tilde #1}
\newcommand\Id{{\mathbb I}}
\newcommand{\cind}[1]{{{}_\lfloor\!\underline{#1\vphantom,}\!{}_\rfloor}}
\newcommand{\Oi}{{\mathtt\OldTheta}}
\renewcommand{\L}{{\mathtt L}}
\newcommand{\pe}{{\ge0}}
\newcommand{\newstuff}{\medskip\centerline{\bf ***}}
\newcommand{\trab}{tr\(_{AB}\)}
\newcommand{\trabh}{tr\(_{AB^\hc}\)}
\newcommand{\Aver}[2]{\overline #1^{#2}}
\newcommand{\cmdh}{\usefont{T1}{cmdh}{m}{n}}

\newcommand\eq[1]{Eq.~(\ref{#1})}
\newcommand\Sec[1]{Sec.~\ref{sec:#1}}
\newcommand\footref[1]{footnote~($^{\ref{#1}}$)}

\title{\cmdh Probabilities, Tensors and Qubits}
\author{\em A. Yu.\ Vlasov}
\date{\framebox{26 April 2001}}
\draftsize
\begin{document}
\maketitle
\newcount\nowhr \nowhr=\time \newcount\nowmn \nowmn=\nowhr
\divide\nowhr by 60 \multiply\nowhr by 60
\advance\nowmn by -\nowhr \divide\nowhr by 60
\def\timestart{\number\nowhr:\ifnum\nowmn<10 0\fi\number\nowmn}
\newcommand{\shortmonth}{\ifcase\month\or
  Jan\or Feb\or Mar\or Apr\or May\or Jun\or
  Jul\or Aug\or Sep\or Oct\or Nov\or Dec\fi}
\def\shortdate{\number\day\ \shortmonth\ \number\year}
\typeout{Job: \jobname.tex,  \number\day-\shortmonth-\number\year,  \timestart}
\makeatletter
\renewcommand{\@oddfoot}{\footnotesize\rlap{\LaTeX: {\tt\jobname.tex}}\hfil%
{\normalsize\thepage}\hfil\llap{\shortdate, \timestart}%
\rlap{\raisebox{\footskip}[0pt][0pt]{~\rule{0.4mm}{\textheight}}}%
\rlap{\sf\quad\qisSPb\quad\quad(comments are welcome)}}
\renewcommand{\@evenfoot}{\@oddfoot}
\global\@specialpagefalse 
\makeatother
\begin{abstract}
 In the paper is discussed complete probabilistic description of quantum
systems with application to multiqubit quantum computations. In simplest
case it is a set of probabilities of transitions to some fixed
set of states. The probabilities in the set may be represented linearly via
coefficients of density matrix and it is very similar with description using
mixed states, but also may give some alternative view on specific properties
of quantum circuits due to possibility of direct comparison with classical
statistical paradigm.
\end{abstract}

\section{Introduction}
\label{sec:intro}

Let us consider $n$-dimensional Hilbert
space $\Hil_n$ and set of $N$ fixed unit vectors $\ket{v_\alpha} \in \Hil_n$
denoted as $\Sv\{v_\alpha:\alpha = 1,\ldots,N\}$ or $\Sv_N(\Hil_n)$. Then for
quantum system in
arbitrary state $\ket{\psi} \in \Hil_n$ there are defined $N$ coefficients:
\begin{equation}
p_\alpha \equiv \Abs{\brkt{v_\alpha}{\psi}}^2.
\label{pureprob}
\end{equation}
The $p_\alpha$ is probability to find $\ket{\psi}$ in state $\ket{v_\alpha}$
due to measurement described by projector:
\begin{equation}
\P_\alpha \equiv \P(v_\alpha) \equiv \vxv{v_\alpha}.
\label{proj}
\end{equation}
For {\em mixed} state with density matrix $\rho$ it is possible to use
instead of \eq{pureprob}:
\begin{equation}
p_\alpha \equiv \bra{v_\alpha}\,\rho\,\ket{v_\alpha} = \Tr(\rho\,\P_\alpha).
\label{mixprob}
\end{equation}
For pure state with $\rho = \vxv{\psi}$ the \eq{mixprob} coincides
with \eq{pureprob}.

Due to \eq{mixprob} probabilities of transitions $p_\alpha(\rho)$ are simply
linear functions of density matrix and for properly chosen set of vectors
$v_\alpha$ using only the coefficients $p_\alpha$ may be equivalent to
description by density matrices. For example in the paper is shown that
for such a set any quantum channel can be described as linear transformation
of $N$-dimensional vector of probabilities, i.e. simply as $N{\times}N$
matrix. For $n$-qubit system it is possible to find a simple set with
$6^n$ vectors, then the probabilities can be ordered in a tensor with $n$
indexes, but only $4^n$ components are linearly independent. It may be
also described by $4^n$ special linearly independent parameters ordered in
4-dimensional real tensor with $n$ indexes. A natural and symmetric
description of any quantum circuits as linear transformations of these
tensors is discussed below in this paper.

\section{General description}
\label{sec:gen}

Let us consider $N$ probabilities $p_\alpha$ \eq{mixprob} as
formal vector $\ve p =(p_1,\ldots,p_N)$. Let $H(n)$ is
space of Hermitian $n{\times}n$ matrices $\mbf{H} \in H(n)$,
$\mbf{H}{}^{lk} = \mbf{\bar H}{}^{kl}$
and $H_{\rm ph}(n)$ is ``physical'' subspace
of {\em nonnegative definite matrices with trace one},
\begin{equation}
H_{\rm ph}(n) = \{\rho ~:~ \rho \in H(n),
\ \Tr \rho = 1, \ \bra{v}\,\rho\,\ket{v} \ge 0, \: \forall\,v \},
\label{Hphys}
\end{equation}
then $\Sv_N$ via
\eq{mixprob} produces linear maps $\L_\Sv \colon H_{\rm ph}(n) \to \R_\pe^N$
or $\L_\Sv \colon H(n) \to \R^N$. Here $H(n)$ is linear space described by
$n^2$ real parameters and $H_{\rm ph}(n)$ is subspace
of physically possible density matrices $\rho \in H_{\rm ph}(n)$,
$\L_\Sv \colon \rho \mapsto \ve p$.

Let us call the set $\Sv_N(\Hil_n)$ {\bf representative} if exists {\em right
inverse} of $\L_\Sv$, i.e., a linear operator $\li\L_\Sv\colon\R^N\to H(n)$
with property:
\begin{equation}
\L_\Sv \circ \li \L_\Sv = \Id, \quad \mbox{i.e.}, \quad
\li \L_\Sv(\L_\Sv \mbf{H}) = \mbf{H}, \quad \forall \mbf{H} \in H(n).
\label{OCompl}
\end{equation}
where symbol ``$\circ$'' is used for composition of operators (left one acts
first).

The space $H(n)$ is described by $n^2$ real parameters and for {\bf minimal
representative} set with $n^2$ vectors $\Sv_{n^2}(\Hil_n)$ the $\L_\Sv$ is
left and right inverse (or, simply, inverse: $\li\L_\Sv = \L_\Sv^{-1}$) and
together with \eq{OCompl} it is possible to write:
\begin{equation}
\li \L_\Sv \circ \L_\Sv = \Id, \quad \mbox{i.e.}, \quad
 \L_\Sv(\li \L_\Sv \ve p) = \ve p, \quad \forall \ve p \in \R^{n^2},
\label{RCompl}
\end{equation}
but if $n^2 > N$ the map $\L_\Sv \circ \li \L_\Sv$ is not identity, it is
projector on some $n^2$-dimensional subspace $\V^{n^2} \subset \R^N$.
The space $\V^{n^2}$ is image of map $\L_\Sv\colon H(n) \to \R^N$.

If representative set is not minimal, the $\li\L_\Sv$ is not unique.
There exists $n^2(N-n^2)$-dimensional space of linear maps
$\li\L_\Sv^\perp$ with property $\li\L_\Sv^\perp(\ve v) = \mbox{\sl 0}$,
$\forall \ve v\in\V$ (here {\sl 0} is zero Hermitian matrix)
and so instead of $\li\L_\Sv$ it is possible to use any other
$\li\L'_\Sv = \li\L_\Sv + \li\L_\Sv^\perp$. Due to such property
results of alternative methods of construction of $\li\L_\Sv$ may
differ on $\li\L_\Sv^\perp$ if set $\Sv$ is not minimal.

\smallskip

For physical space of density matrices \eq{Hphys} due to condition
$\Tr \rho = 1$, it is possible to express one of component using
other with some affine relation. It is possible to introduce
some {\bf affine minimal} set with $n^2-1$ elements, but instead of
linear equation \eq{OCompl} we will have affine equation:
\begin{equation}
\rho = \li\mathtt{A}_\Sv(\L_\Sv \rho) + \li\mathtt{H}_0 ,
\quad \forall \rho \in H_{\rm ph}(n).
\label{ACompl}
\end{equation}

\smallskip

Let us call a representative set $\Sv_N$ {\bf complete}, if it can be
expressed as union of few orthonormal bases of $\Hil_n$, i.e., for {\em any}
$v_\alpha \in \L_\Sv$ it is possible to find other $n-1$ vectors from $\L_\Sv$
to form an orthogonal basis for $\Hil_n$.

Let us call the complete set $\Sv_N$ {\bf almost perfect}, if it can be
expressed as {\em disjoint} union of few orthonormal bases of $\Hil_n$
and {\bf perfect}, if for any $v_\alpha \in \L_\Sv$ there is
{\em unique} choice of $n-1$ vectors from $\L_\Sv$ to form the
orthogonal basis for $\Hil_n$.

The definitions of complete, almost perfect and perfect sets are useful
for consideration of elementary measurements with $n$ possible outcomes.

It is clear from definition that {\em perfect} set is also {\em almost
perfect}. The example of {\em almost perfect} set that is {\em not perfect},
is one of main themes of this paper. It is complete set for $m\ge2$ qubits
introduced in \Sec{qubits}. The set is almost perfect, but not perfect,
due to counterexample in \Sec{twoq} (page \pageref{sec:twoq}).
It will be shown below (page \pageref{notmin}) that almost perfect set
may not be minimal.

\newstuff

Let $\ket{1},\ldots,\ket{n}$ is a basis of $\Hil_n$ and $\rho_{kl}$ are
components of density matrix $\rho = \sum_{k,l=1}^n{\rho_{kl}\ket{k}\bra{l}}$
and the map $\L_\Sv\colon \rho \mapsto \ve p$ is defined as above by \eq{mixprob}.
Then the definition of $\li\L_\Sv$ using
\eq{OCompl} corresponds to existence of $n^2 N$ complex coefficients
$c_{kl}^\alpha$ (some of them may be zero and the choice is unique only for
{\em minimal} set), where $k,l = 1,\ldots,n$ and $\alpha = 1,\ldots,N$ with
property:
\begin{equation}
 \rho_{kl} = \sum_{\alpha=1}^N {c_{kl}^\alpha p_\alpha}
\quad \mbox{(there is no summation on $k,l$) .}
\label{klCompl}
\end{equation}

Let us introduce a basis on space $\C^{n\times n}$ of all complex matrices.
Elements of the basis are $\mbf{E_{kl}} = \ket{k}\bra{l}$, i.e., matrices with
only one unit in position $(k,l)$:
${(\mbf{E_{kl}})}_{ij} = \delta_{ki}\delta_{lj}$.
Let us show, {\em that a set is representative and \eq{klCompl} is true
for some coefficients $c_{kl}^\alpha$ if and only if it is possible to
express all $n^2$ elements of the basis as linear combinations of projectors
$\P_\alpha$}:
\begin{equation}
 \mbf{E_{lk}} = \ket{l}\bra{k} = \sum_{\alpha=1}^N {c_{kl}^\alpha \P_\alpha}
\quad \mbox{(there is no summation on $k,l$) .}
\label{Elk}
\end{equation}

First, \eq{klCompl} is direct sequence of \eq{Elk} because:
\[
\sum_{\alpha=1}^N {c_{kl}^\alpha p_\alpha} =
\sum_{\alpha=1}^N {c_{kl}^\alpha \Tr(\rho\,\P_\alpha)} =
\Tr{\Bigl(\rho \sum_{\alpha=1}^N c_{kl}^\alpha \P_\alpha\Bigr)} =
\Tr\BIG({\rho\,\ket{l}\bra{k}}) = \bra{k}\,\rho\,\ket{l} = \rho_{kl}
\]

Now let us show that for representative set always exists some decomposition
\eq{Elk}. It maybe not obvious, the \eq{Elk} looks too general,
because we considered basis of whole matrix algebra $\C^{n\times n}$ with
$n^2$ complex (i.e. $2n^2$ real) parameters instead of space of Hermitian
matrices $H(n)$ described by only $n^2$ real parameters.

The condition for representative set is existence of map $\li \L_\Sv$, such
that composition $\L_\Sv \circ \li\L_\Sv$ is identity $H(n) \to H(n)$
in diagram:
\begin{equation}
 H(n) \stackrel{\L_\Sv}{\longrightarrow} \R^N
      \stackrel{\li\L_\Sv}{\longrightarrow} H(n),
\quad \L_\Sv \circ \li\L_\Sv = \Id.
\label{Hn2Hn}
\end{equation}
It is necessary and enough to satisfy the identity
for whole basis of $H(n)$ with $n^2$ elements. It is possible if $N\ge n^2$
and in this case image of $\L_\Sv$ must be some $n^2$-dimensional
subspace $\V^{n^2} \subset \R^N$.

On the other hand, the \eq{Elk} is equivalent with extension of the
$\li \L_\Sv$, $\L_\Sv$ and identity of $\L_\Sv \circ \li\L_\Sv$ on whole
space $\C^{n\times n}$:
\begin{equation}
 \C^{n\times n} \stackrel{\L^\C_\Sv}{\longrightarrow} \C^N
                \stackrel{\li \L^\C_\Sv}{\longrightarrow} \C^{n\times n},
\quad \L^\C_\Sv \circ \li\L^\C_\Sv = \Id.
\label{Cnn2Cnn}
\end{equation}
But let us show that identity of $\L_\Sv \circ \li\L_\Sv$ for basis
elements $H(n)$ described in \eq{Hn2Hn} is enough for identity of the
map for basis of $\C^{n\times n}$, i.e., the \eq{Cnn2Cnn} is not more
general.

A basis of $H(n)$ is $n(n+1)/2$ matrices
\begin{equation}
\mbf{H^+_{kl}} = (\mbf{E_{kl}}+\mbf{E_{lk}})/2,\quad k \ge l
\label{Hps}
\end{equation}
(i.e. $\mbf{H^+_{kk}} = \mbf{E_{kk}}$) and $n(n-1)/2$ matrices
\begin{equation}
\mbf{H^-_{kl}} = i(\mbf{E_{kl}}-\mbf{E_{lk}})/2,\quad k > l.
\label{Hms}
\end{equation}
But if some linear map $\C^{n\times n} \to \C^{n\times n}$ is identity on
these elements, it is also identity on linear combinations, but any
matrix of basis $\mbf{E_{kl}}$ of $\C^{n\times n}$ may be expressed
via $\mbf{H^+_{kl}}$ and $\mbf{H^-_{kl}}$:
\begin{equation}
\mbf{E_{kl}} = \left\{
\begin{array}{cl}
\mbf{H^+_{kl}} - i \mbf{H^-_{kl}},& k > l \\
\mbf{H^+_{kk}},& k = l \\
\mbf{H^+_{lk}} + i \mbf{H^-_{lk}},& k < l \\
\end{array}
\right.
\label{EHpm}
\end{equation}

So $\L^\C_\Sv \circ \li\L^\C_\Sv$ is identity and \eq{Elk} is necessary
and enough, it is simply expression for $\li \L^\C_\Sv$ written in
basis $\mbf{E_{lk}}$.

\smallskip

The following {\bf proposition} summarizes the consideration above.
\\{\em Three properties are equivalent:}
\begin{enumerate}
 \item\label{R1}
  Set $\Sv(v_\alpha)$ is representative ($\li\L_\Sv$ exists).
 \item\label{R2}
  Any complex matrix may be represented as linear combination
 of projectors $\P(v_\alpha)$ with {\em complex coefficients} --- it
 is enough to show for any basis, for example $\mbf{E_{kl}}$, see \eq{Elk}.
 \item\label{R3}
  Any Hermitian matrix may be represented as linear combination
 of projectors $\P(v_\alpha)$ with {\em real coefficients} --- it
 is enough to show for any basis, for example $\mbf{H^\pm_{kl}}$ above.
\end{enumerate}

For minimal representative set the construction of $\li\L_\Sv$
(item \ref{R1} in the proposition above) and both decompositions
(items \ref{R2} and \ref{R3}) are unique.

\smallskip

Now it is possible to show that
{\em \label{notmin} (almost) perfect set may not be minimal.}
For $n$-dimensional Hilbert space minimal set must have only $n^2$ elements.
If it could be (almost) perfect set, it can be considered as disjoint
union of $n$ bases $\Sv\{v_{\alpha(k,j)}\}$, where $k = 1,\ldots,n$ is label
of basis and $j = 1,\ldots,n$ is number of element in $k$-th basis.
But in this case we would have $n$ decompositions of same matrix (unit)
$\Id = \sum_{j=1}^n\P_{\alpha(k,j)}$, $\forall k$ and so it may be no more
than $n^2-n+1$ different linearly independent combinations of the projectors
$\P_{\alpha(k,j)}$ instead of $n^2$ and the set may not be representative.

\smallskip

The complex basis $\mbf{E_{kl}}$ is ``\trabh-orthonormal'', i.e., orthonormal
with respect to scalar product on space of matrices defined as
\begin{equation}
 \langle\!\langle A , B \rangle\!\rangle_\hc \equiv \Tr(A\,B^\hc).
\label{ortmat}
\end{equation}
Due to it there is simple relation between expression for the basis \eq{Elk}
and an expansion \eq{klCompl} of $\li \L$ for representative set.
For Hermitian matrices \eq{ortmat} is close related with \eq{mixprob}, because
two different scalar products coincide:
\begin{equation}
 \langle\!\langle A , B \rangle\!\rangle \equiv \Tr(A\,B) =
 \Tr(A\,B^\hc) \equiv \langle\!\langle A , B \rangle\!\rangle_\hc.
\label{ortHerm}
\end{equation}
The basis of Hermitian matrices defined by \eq{Hps} and \eq{Hms} is not
\trab-orthonormal, but it is possible to use standard Gram-Schmidt procedure
of orthogonalization for given norm (cf. Ref.~\cite{Pavia}). An example of
\trab-orthonormal Hermitian basis for qubits with dimension of Hilbert space
is $2^n$ will be discussed later.

\newstuff

Other application of $\mbf{E_{kl}}$ is following theorem.

{\bf Composition Theorem}: {\em Let $\Hil_n^{(1)}$ and $\Hil_m^{(2)}$ are two
Hilbert spaces with dimensions $n$ and $m$ and
$\Sv_N\{v_\alpha \in \Hil_n^{(1)}\}$ and $\Sv_M\{u_\beta \in \Hil_m^{(2)}\}$
are representative (complete, almost perfect) sets with $N$ and $M$ vectors.
Then set of $NM$ vectors
\mbox{$\Sv_{NM}\{v_\alpha \ox u_\beta \in \Hil_n^{(1)}\ox\Hil_m^{(2)}\}$}
is representative (complete, almost perfect) set for composite system.}

Let us prove the theorem for representative sets, because
for complete sets it is trivial implication.
For a proof it is enough to use condition \eq{Elk} for $\Hil_n^{(1)}$ and
$\Hil_m^{(2)}$:
\[
\ket{l_1}\bra{k_1} =
\sum_{\alpha=1}^N {c_{k_1 l_1}^{_{(1)}\alpha} \P(v_\alpha)},
\quad
\ket{l_2}\bra{k_2} =
\sum_{\beta=1}^M {c_{k_2 l_2}^{_{(2)}\beta} \P(u_\beta)},
\]
(where $\ket{l_1},\ket{k_1},\ket{v_\alpha} \in \Hil^{(1)}$ and
$\ket{l_2},\ket{k_2},\ket{u_\beta} \in \Hil^{(2)}$)
together with properties:
\[
\P(v \ox u) = \P(v) \ox \P(u),
\quad
\ket{l_1 l_2}\bra{k_1 k_2} = \ket{l_1}\bra{k_1}\otimes\ket{l_2}\bra{k_2}.
\]
So for $n^2 m^2$ elements of basis in space of all matrices
$\C^{nm\times nm} \cong \C^{n\times n}\ox\C^{m\times m}$ we have:
\begin{equation}
\ket{l_1 l_2}\bra{k_1 k_2} =
\sum_{\alpha=1}^N\sum_{\beta=1}^M
{c_{k_1 l_1}^{_{(1)}\alpha}c_{k_2 l_2}^{_{(2)}\beta}\P(v_\alpha \ox u_\beta)}.
\label{E1E2}
\end{equation}

The \eq{E1E2} corresponds to \eq{Elk} for $nm$-dimensional Hilbert space
$\Hil_{nm}^{(1,2)} = \Hil_n^{(1)} \ox \Hil_m^{(2)}$ with representative set
$\Sv_{NM}(\Hil^{(1,2)})$ with $NM$ vectors, if to use ``compound indexes''
like $\cind{k_1 k_2} \leftrightarrow (k_1 - 1) m + k_2$ and
$\cind{\alpha \beta} \leftrightarrow (\alpha - 1) M + \beta$:
\begin{equation}
\mbf{E_{\cind{l_1 l_2}\cind{k_1 k_2}}} = \sum_{\cind{\alpha\beta} = 1}^{NM}%
{c_{\cind{k_1 k_2}\cind{l_1 l_2}}^{\cind{\alpha\beta}}
\P_{\cind{\alpha\beta}}},
\label{E12}
\end{equation}
where $c_{\cind{k_1 k_2}\cind{l_1 l_2}}^{\cind{\alpha\beta}} =
c_{k_1 l_1}^{_{(1)}\alpha}c_{k_2 l_2}^{_{(2)}\beta}$ and
$\P_{\cind{\alpha\beta}} = \P(v_\alpha \ox u_\beta)$.

\smallskip
A proof that tensor product of complete (almost perfect) sets is complete
(almost perfect) is directly implied because tensor product of two bases
is basis. Tensor product of perfect set is almost perfect, but not necessary
perfect, it follows from counterexample in \Sec{twoq}.

\medskip

It is also possible to suggest proof without complexification,
because tensor product of Hermitian matrices is also Hermitian,
but it is not discussed here due to rather technical points.
Such approach may be more clear from consideration with Pauli
matrices in \Sec{qubits}.

\section{Some applications of representative sets of projectors}
\label{sec:appl}

The \eq{klCompl} shows that for representative set using of vector $\ve p$
is equivalent to description of quantum system by density matrix.

\paragraph{Example 1:} For any unitary operator $U$ on $\Hil$,
$\ket{\psi'} = U \ket{\psi}$ and
\begin{equation}
 \rho' = U \rho\, U^\hc
\label{urhou}
\end{equation}
there exists $N{\times}N$ matrix $A_U$, $\ve p' = A_U \ve p$,
there $\ve p' \equiv \ve p_{\psi'}$ (here $A_U$ does not depend on
$\ket{\psi}$). Really, the operator $U$ induces a linear transformation
$U\ox U^\hc$ \eq{urhou} on space of Hermitian matrices. Let us denote it as
$\mathtt{U^{1,1}} \colon \rho \mapsto \rho'$.
Then $A_U = \li\L_\Sv \circ \mathtt{U^{1,1}} \circ \L_\Sv$ as follows from
diagram\footnote{See also \footref{hist} below on page \pageref{hist}}:
\begin{equation}
\setlength{\arraycolsep}{0.1em}
\renewcommand{\arraystretch}{1.5}
 \begin{array}{ccc}
   \ve p & \stackrel{\textstyle A_U}{\To} & \ve p' \\
    \llap{$\li{\L_\Sv}$}\Bigl\downarrow\!\!\Bigr\uparrow\rlap{$\L_\Sv$} &&
    \Big\uparrow {\lefteqn{\L_\Sv}} \\
   \rho & \smash{\stackrel{\textstyle\mathtt{U^{1,1}}}{\To}} & \rho'
 \end{array}
\label{comdiag}
\end{equation}

\medskip

Let us consider construction of matrix $A_U$ more directly. Components
of the vector $\ve p'$ are written as:
\begin{equation}
 p'_\alpha = \Tr(U\rho\,U^\hc \P_\alpha) = \Tr(\rho\,U^\hc \P_\alpha U)
\end{equation}
and because due to third item of proposition in \Sec{gen} any unitary matrix
can be represented as linear combination of $\P_\alpha$, we have
\begin{equation}
 U^\hc \P_\alpha U = \sum_{\beta=1}^N A_\alpha^\beta \P_\beta,
\end{equation}
where for given $\alpha$, $A_\alpha^1,\ldots,A_\alpha^N$ are real
coefficients of decomposition of Hermitian matrix $U^\hc \P_\alpha U$.
The coefficients may be unique only if $N=n^2$ and $\P_\beta$ are linearly
independent, i.e., for minimal representative set. On the other hand,
$A_\alpha^\beta$ are components of desired matrix $A_U$,
$\ve p' = A_U \ve p$, because
\begin{equation}
 p'_\alpha = \Tr(\rho \sum_{\beta=1}^N A_\alpha^\beta \P_\beta)
= \sum_{\beta=1}^N A_\alpha^\beta \Tr(\rho \P_\beta)
= \sum_{\beta=1}^N A_\alpha^\beta p_\beta.
\end{equation}

If set is not minimal, the decomposition is not unique, for example
it is possible to use only $n^2$ linearly independent projectors
$\P_\alpha$ between $N > n^2$
and already this choice is not unique. There is a problem
with calculation of coefficients $A_\alpha^\beta$ even for minimal set,
but it is convenient for set of matrix orthogonal in respect of
some norm. The complex extension with matrices $\mbf{E_{kl}}$ orthogonal in
norm \eq{ortmat} was discussed already and for system ($n=2^m$) of $m$-qubits
basis of Hermitian matrix obtained from $4^m$ different products of Pauli
matrices orthogonal in both norms \eq{ortHerm} and \eq{ortmat} is used
below \eq{H2nbas}.

Let us suggest, that we choose some basis of $n^2$ Hermitian matrices
$\mbf{H_K}$ and express all matrices $\P_\alpha$,
\begin{equation}
 \P_\alpha = \sum_{K=1}^{n^2} h^K_\alpha \mbf{H_K}.
\end{equation}
The coefficients $h^K_\alpha$ are unique and if basis $\mbf{H_K}$ is
\trab-orthonormal
\begin{equation}
\Tr(\mbf{H_J \, H_K}) = \delta\mbf{_{JK}}
\end{equation}
then the coefficients may be expressed as
\begin{equation}
 h^\alpha_K = \Tr(\P_\alpha \mbf{H_K})
\end{equation}

Let us introduce new parameters:
\begin{equation}
 \tilde p_K = \Tr(\rho \mbf{H_K}).
\label{parHbas}
\end{equation}
Then it is possible to express all $N \ge n^2$ probabilities $p_\alpha$
using the parameters:
\begin{equation}
p_\alpha = \Tr(\rho\,\P_\alpha) =
\Tr(\rho\sum_{K=1}^{n^2} h^K_\alpha \mbf{H_K}) =
\sum_{K=1}^{n^2} h^K_\alpha \Tr(\rho\mbf{H_K}) =
\sum_{K=1}^{n^2} h^K_\alpha \tilde p_K.
\label{tp2p}
\end{equation}

If set is not minimal, coefficients $\tilde h^\alpha_K$ of inverse
transformation:
\begin{equation}
\mbf{H_K} = \sum_{\alpha=1}^N \tilde h^\alpha_K \P_\alpha,
\end{equation}
and
\begin{equation}
\tilde p_K = \sum_{\alpha=1}^N \tilde h^\alpha_K p_\alpha
\end{equation}
are not unique.

The parameters $\tilde p_K$ can be considered as a $n^2$-dimensional vector
$\tilde\ve p$, but now transformation $\tilde\ve p' = \tilde A_U\tilde\ve p$
always unique and can be calculated directly, let
\begin{equation}
 \tilde p'_K = \Tr(U\rho\,U^\hc \mbf{H_K}) = \Tr(\rho\, U^\hc \mbf{H_K} U),
\label{UrUH}
\end{equation}
it is possible to use decomposition
\begin{equation}
 U^\hc \mbf{H_K} U = \sum_{J=1}^{n^2} \tilde A_K^J \mbf{H_J},
\end{equation}
and so
\begin{equation}
 \tilde p'_K = \sum_{J=1}^{n^2} \tilde A_K^J \tilde p_J,
\end{equation}
where matrix $\tilde A_U$ is expressed as:
\begin{equation}
(\tilde A_U){}_K^J = \Tr(\mbf{H_J} U^\hc \mbf{H_K} U).
\label{tAKJ}
\end{equation}

Initial matrix $A_U$ can be expressed from $\tilde A_U$ using linear maps
$h^K_\alpha$ and $\tilde h^\alpha_K$ between $p_\alpha$ and $\tilde p_K$.
If set is not minimal, $A_U$ depends on $\tilde h^\alpha_K$ and is not
unique. It is unique only restriction of $A_U$ on $n^2$-dimensional linear
subspace $\V$ discussed above. It is possible also directly work with
$\tilde p_K$, because all probabilities can be expressed using the parameters
and expressions \eq{tp2p} where $h^K_\alpha$ are always unique. The
$\tilde p_K$ can be considered as ``coordinates'' on $\V$.

\medskip

The idea to write transformations of state of quantum system in terms
of probabilities (weights) was also suggested in relation with some other
problems in \cite{GZ90,GZ92,BG}, but authors found {\em linear}
transformation only for $n=2$ and work only with pure states.\footnote{%
\label{hist}Initial version of present paper (May 2000) appears as positive
answer on question if the maps like $A_U$ may be linear for $n>2$. It should
be mentioned also, that for pure state due to polynomial relations between
coefficients of density matrix like $\rho_{kl}\rho_{mn}=\rho_{kn}\rho_{ml}$
it may be possible to express some probabilities in minimal set from the
others by some nonlinear functions and so use nonlinear transformations
with lesser amount of vectors $v_\alpha$ instead of $\li\L$ and $A_U$,
but it is not discussed here. On the other hand, even for linear $A_U$ the
existence of inverse map $\li\L_\Sv$ for $\L_\Sv$ is too strong requirement.
If to look {\em only} for the linear map $A_U$, it is enough to consider
condition that diagram \eq{comdiag} is commutative. In this case it is
reasonable to consider irreducible representations of tensor product
$U(n)\times U^\hc(n)$. It can be shown, that such product is sum of two
irreps with dimensions are $n^2-1$ and $1$. The second one is simply scalar
representation corresponding to constancy of density matrix trace and so
splitting off the one-dimensional irrep may be explained without applications
of theory of group representations. Due to such situation, instead of
commutativity of diagram \eq{comdiag} here was used straightforward, but less
general suggestion about existence of $\li\L_\Sv$, especially because the map
is anyway necessary for other applications discussed in this paper.}

\paragraph{Example 2:} The representation may be even more convenient for
general quantum channel \cite{chann}:
\begin{equation}
 \rho' = \sum_k V_k \rho V_k^\hc,
 \quad \sum_k V_k V_k^\hc = \Id.
\label{superho}
\end{equation}
In this case linear transformation\footnote{Sometimes it is called
``superoperator.''} $\mathtt V\colon\rho'\mapsto\rho$ is
defined by some set $\{V_k\}$ and the set is even not unique, but because
the diagram \eq{comdiag} is valid for any linear map $H(n) \to H(n)$,
any quantum channel can be expressed using only one matrix
$A_{\mathtt V} = \sum_k{A_{V_k}}$, $\ve p' = A_{\mathtt V} \ve p$.
For minimal set matrix $A_{\mathtt V}$ is unique,
overwise unique only restriction of $A_{\mathtt V}$ on $n^2$-dimensional
linear subspace $\V$.

Applications of parameters $\tilde \ve p$ here is also justified. It is
directly followed from consideration of previous example. A matrix of map
$\tilde A_{\mathtt V}\colon \tilde \ve p \mapsto \tilde \ve p'$
is also always unique.

\paragraph{Example 3:} Any expressions for probabilities can be found using
only elements of vector $\ve p$ (cf. Ref.~\cite{GZ90}).  For any
vector $v \in \Hil$ there is linear map $B_v\colon \R^N{\to}\R$, defined as
\begin{equation}
\bra{v}\,\rho\,\ket{v} = B_v\!\ve p = \bra{v}\li \L_\Sv\ve p\ket{v}
\label{Bdef}
\end{equation}
($B_v$ depends only on $v$, not $\ket{\psi}$). Let $\ve p_\psi$ corresponds
to some pure state $\psi$, $\ve p_\psi = \L_\Sv\BIG({\vxv{\psi}})$
by definition of $\L_Sv$ via \eq{pureprob}, \eq{mixprob}, then \eq{Bdef}
can be rewritten as $\Abs{\brkt{v}{\psi}}^2 = B_v\!\ve p_\psi$.
It is probability of transition. Let $\ve p_v = \L_\Sv\BIG({\vxv{v}})$, then
because $\Abs{\brkt{v}{\psi}}^2 = \Tr\BIG({\vxv{v}\,\vxv{\psi}}) =
\Tr{\bigl(\li \L_\Sv(\ve p_v) \li \L_\Sv(\ve p_\psi)\bigr)}$ is some bilinear
map $\R^N \times \R^N \to \R$, due to general theorem of linear algebra, it
is scalar product $(\ve p_v , \ve p_\psi)_\Sv = \Abs{\brkt{v}{\psi}}^2$
defined by some matrix $G_\Sv$:
\begin{equation}
(\ve p , \ve q)_\Sv =
\sum_{\alpha,\beta = 1}^N {G_\Sv^{\alpha\beta} p_\alpha q_\beta}.
\label{metG}
\end{equation}
So it is possible to define some metric $G_\Sv$ on $\R^N$ with property
that probability of transition for pure states is scalar product in this
metric.

\paragraph{Example 4:} It is possible to calculate average value of any
Hermitian operator $X$ for some observable:
$\Aver X\psi = \bra{\psi}\,X\,\ket{\psi} = \Tr(\li \L_\Sv(\ve p_\psi) X)$
(cf. Refs.~\cite{GZ90}, \cite[\S8.1.1]{GR99}) for pure state.  For mixed
state it is possible to use formula
$\Aver X\rho = \Tr(\rho\,X) = \Tr(\li \L_\Sv(\ve p_\rho) X)$ and if to
represent a Hermitian operator $X$ as some vector
$\ve p_X \equiv \L_\Sv(X) \in \R^N$, then average value of operator $X$ for
some mixed state $\rho$ is expressed via the same metric $G_\Sv$ \eq{metG}
introduced in example above

\begin{equation}
 \Aver X\rho = (\ve p_X,\ve p_\rho)_\Sv .
\end{equation}

\section{Example of representative and complete sets}
\label{sec:anyN}

Let us consider example with Hilbert space $\Hil_n$ and minimal
representative set with $N = n^2$ elements. The $n^2$ vectors are expressed
using $n$ basis vectors $\ket{\alpha}$ as following three families of
vectors:

\begin{tabular}{ll}
$n$ basis vectors itself:& $\ket{v^z_{\alpha}}=\ket{\alpha}$, \\
$\frac{n(n-1)}{2}$ vectors:& $\ket{v^x_{{\alpha\beta}}}=
\frac{1}{\sqrt2}(\ket{\alpha}+\ket{\beta})$, $\alpha < \beta$,\\
$\frac{n(n-1)}{2}$ vectors:& $\ket{v^y_{{\alpha\beta}}}=
\frac{1}{\sqrt2}(\ket{\alpha}+i\ket{\beta})$, $\alpha < \beta$.
\end{tabular}

The set is known also due to application for construction of POVM \cite{CFS}.

The weight vector $\ve p$ with $N=n^2$ real components is composed as
set of numbers from three different families:
\begin{eqnarray*}
p^z_\alpha &\!=\!& \bra{\alpha}\,\rho\,\ket{\alpha} = \rho_{\alpha\alpha}
\\
p^x_{\alpha\beta}
 &\!=\!& \half(\bra{\alpha}+\bra{\beta})\,\rho\,(\ket{\alpha}+\ket{\beta})
=  \half\BIG({\rho_{\alpha\alpha}+\rho_{\beta\beta} +
   (\rho_{\alpha\beta}+\rho_{\beta\alpha})})
\\
p^y_{\alpha\beta}
 &\!=\!& \half(\bra{\alpha}-i\bra{\beta})\,\rho\,(\ket{\alpha}+i\ket{\beta})
= \half\BIG({\rho_{\alpha\alpha}+\rho_{\beta\beta}
   +i(\rho_{\alpha\beta}-\rho_{\beta\alpha})})
\end{eqnarray*}

So $\rho_{\alpha\alpha} = p^z_\alpha$ and
$(\rho_{\alpha\beta}+\rho_{\beta\alpha})/2 = p^x_{\alpha\beta}
- \half(p^z_\alpha+p^z_\beta)$,
$i(\rho_{\alpha\beta}+\rho_{\beta\alpha})/2 = p^y_{\alpha\beta}
- \half(p^z_\alpha+p^z_\beta)$ for $\alpha{<}\beta$
and it is possible to write for arbitrary matrix:
\begin{equation}
 \rho_{\alpha\beta} = \left\{
 \begin{array}{ll}
   p^z_\alpha;& \alpha = \beta \\
   p^x_{\alpha\beta}-i\,p^y_{\alpha\beta}-
   \frac{1+i}{2}(p^z_\alpha+p^z_\beta);
   & \alpha < \beta \\
   p^x_{\beta\alpha}+i\,p^y_{\beta\alpha}-
   \frac{1-i}{2}(p^z_{\alpha}+p^z_{\beta});
   & \alpha > \beta
 \end{array}
 \right.
\label{invL}
\end{equation}
or for Hermitian matrix it is convenient also to use expressions with
$n^2$ real parameters for real and imaginary components of elements
$\rho_{\alpha\beta}$ (here $\alpha < \beta$):
\begin{eqnarray*}
\rho_{\alpha\alpha} &=& p^z_\alpha, \\
\Re\rho_{\alpha\beta} = \Re\rho_{\beta\alpha} &=&
 p^x_{\alpha\beta} - \half(p^z_\alpha+p^z_\beta), \\
\Im\rho_{\alpha\beta}= -\Im\rho_{\beta\alpha} &=&
 p^y_{\alpha\beta} - \half(p^z_\alpha+p^z_\beta).
\end{eqnarray*}

The \eq{invL} is direct construction of linear map $\li \L$ discussed above
and so $n^2$ vectors $v^{\{z,x,y\}}_{\{\alpha,\alpha\beta\}}$ are
representative set.

But the set is not complete. It is possible to add two other
families of vectors for completion:

\begin{tabular}{ll}
$\frac{n(n-1)}{2}$ vectors:& $\ket{v'^x_{{\alpha\beta}}}=
\frac{1}{\sqrt2}(\ket{\alpha}-\ket{\beta})$, $\alpha < \beta$,\\
$\frac{n(n-1)}{2}$ vectors:& $\ket{v'^y_{{\alpha\beta}}}=
\frac{1}{\sqrt2}(\ket{\alpha}-i\ket{\beta})$, $\alpha < \beta$.
\end{tabular}

\section{Complete sets for quantum circuits}
\label{sec:qubits}

\subsection{One qubit}
\label{sec:oneq}

The complete set described in the previous example has $2n^2-n$ elements
and for case of qubit $n=2$ with basis $\ket{0},\ket{1} \in \Hil_2$ it is
especially simple and symmetric {\em perfect} set with six elements:
\[
\begin{array}{ll}
\ket{0^z} \equiv \ket{v^z_1} = \ket{0}, &
\ket{1^z} \equiv \ket{v^z_2} = \ket{1},\\
\ket{0^x} \equiv \ket{v^x_{12}} = \frac{1}{\sqrt2}(\ket{0}+\ket{1}), &
\ket{1^x} \equiv \ket{v'^x_{12}}= \frac{1}{\sqrt2}(\ket{0}-\ket{1}), \\
\ket{0^y} \equiv \ket{v^y_{12}} = \frac{1}{\sqrt2}(\ket{0}+i\ket{1}), &
\ket{1^y} \equiv \ket{v'^y_{12}}= \frac{1}{\sqrt2}(\ket{0}-i\ket{1}).
\end{array}
\]

It is simple to check that they are eigenvectors of Pauli matrices with
eigenvalues $\pm 1$:
\begin{equation}
\sigma^\mu\ket{\Oi^\mu} = \lambda_\Oi\ket{\Oi^\mu}
\quad \mbox{(no summation)},
\quad \lambda_\Oi = (-1)^\Oi,
\label{eiqubit}
\end{equation}
where $\Oi = 0,1$ and $\mu \in \{x,y,z\}$.
The \eq{eiqubit} can be simply checked using expression for projectors:
\begin{equation}
\P_\Oi^\mu \equiv \P(\Oi^\mu) = \vxv{\Oi^\mu} =
\half(\Id + \lambda_\Oi \, \sigma^\mu),
\end{equation}
\begin{equation}
\sigma^\mu = \lambda_\Oi(2\P_\Oi^\mu - \Id) =
2\P_0^\mu - \Id = \Id - 2\P_1^\mu = \P_0^\mu - \P_1^\mu,
\quad \P_0^\mu + \P_1^\mu = \Id,
\end{equation}
then
\[
\sigma^\mu\ket{\Oi^\mu} = \lambda_\Oi(2\P_\Oi^\mu - \Id)\ket{\Oi^\mu}
= \lambda_\Oi(2\P_\Oi^\mu\ket{\Oi^\mu} - \ket{\Oi^\mu})
= \lambda_\Oi(2\ket{\Oi^\mu} - \ket{\Oi^\mu})
= \lambda_\Oi\ket{\Oi^\mu}.
\]

Because operator of spin can be expressed\footnote{in Planck's units} as
$\half\sigma^\mu$ the \eq{eiqubit} shows that $\ket{0^\mu}$ and $\ket{1^\mu}$
correspond to spin $+\half$ or $-\half$ respectively for measurements in three
orthogonal directions $\mu \in \{x,y,z\}$.

Six equations for probabilities $p_0^x,\ldots,p_1^z$ are:
\begin{equation}
p^\mu_\Oi = \bra{\Oi^\mu}\,\rho\,\ket{\Oi^\mu} = \Tr(\rho\,\P_\Oi^\mu)
= \Tr(\rho\,\half(\Id + \lambda_\Oi \, \sigma^\mu))
= \half\BIG({1 + \lambda_\Oi \Tr(\rho\, \sigma^\mu)}).
\label{densig}
\end{equation}

Let us now use together with $\mu\in\{x,y,z\}$ numerical indexes $\mu = 1,2,3$
and also $\mu,\nu = 0,\ldots,3$:
\[
\sigma^0 \equiv \Id, \quad \sigma^1 \equiv \sigma^x,
\quad \sigma^2 \equiv \sigma^y, \quad \sigma^3 \equiv \sigma^z.
\]
It is a basis of space of Hermitian matrices:
\begin{equation}
\mbf{H} = \sum_{\nu=0}^3 a_\nu \sigma^\nu,
\quad \mbf{H} \in H(2),\quad a_\nu \in \R.
\label{H2bas}
\end{equation}
Because of a property of the Pauli matrices:
\begin{equation}
\Tr(\sigma^\mu \sigma^\nu) = 2\delta_{\mu\nu}
\end{equation}
they form the \trab-orthogonal basis and
it is simple to find the coefficients $a_\nu$ in \eq{H2bas}:
\begin{equation}
a_\nu = \half \Tr(\mbf{H}\,\sigma^\nu).
\label{asigm}
\end{equation}

For density matrix it is always $a_0=\half$, but here is used all four
indexes due to application to composition below. Let us denote
$\tilde p_\mu = p_0^\mu - p_1^\mu$ and $\tilde p_0 = p_0^\mu + p_1^\mu = 1$,
$\forall \mu = 1,2,3$. Then from \eq{densig} follows:
\begin{equation}
 \tilde p_\nu = \Tr(\rho\,\sigma^\nu),\quad \nu = 0,\ldots,3.
\label{psigm}
\end{equation}
Up to insignificant multiplier $\half$, $\tilde p_\nu$ correspond to
parameters $a_\nu$ just defined in \eq{asigm} or $\tilde p_K$ introduced
earlier in \eq{parHbas}. The density matrix can be expressed as
\begin{equation}
\rho = \half \sum_{\nu=0}^3 \tilde p_\nu \sigma^\nu,
\label{pH2bas}
\end{equation}
due to \eq{H2bas} and probabilities $p^\mu_\Oi$ can be expressed as
\begin{equation}
p^\mu_\Oi = \half(\tilde p_0 + \lambda_\Oi \, \tilde p_\mu),
\quad \mu = 1,2,3, \quad \Oi = 0,1, \quad \tilde p_0 = 1
\end{equation}
due to \eq{densig}.

A simple transformation property of parameters $\tilde p_\mu$
(here $\mu = 1,2,3$)
for unitary 1-qubit gate (cf. example 1 in \Sec{appl} and \cite{BG}) is
related with 2-1 homomorphism $SO(3)$ and $SU(2)$ groups --- for any matrix
$U \in SU(2)$ there is $3{\times}3$ matrix $O \in SO(3)$ with property:
\begin{equation}
U \sigma^\mu U^{-1} = \sum_{\nu=1}^3{O_{\mu\nu} \sigma^\nu}.
\end{equation}
For a map $A_U\colon \tilde p_\mu \mapsto \tilde p'_\mu$ we have:
\[
\tilde p'_\mu = \Tr(U\rho\,U^\hc \sigma^\mu) = \Tr(\rho\,U^\hc \sigma^\mu U)
= \Tr\Bigl(\rho\sum_{\nu=1}^3{O^{-1}_{\mu\nu} \sigma^\nu}\Bigr)
= \sum_{\nu=1}^3{O^{-1}_{\mu\nu} \tilde p_\nu}.
\]
So for $\tilde \ve p_\mu \equiv (\tilde p_1,\tilde p_2,\tilde p_3)$
it is possible simply to write:
\begin{equation}
 \tilde \ve p' = O_U^{-1} \tilde \ve p, \quad O_U \in SO(3).
\label{qurot}
\end{equation}
The \eq{qurot} shows that one-qubit gates corresponds to 3D rotations
of vector $(\tilde p_1,\tilde p_2,\tilde p_3)$ \cite{BG} and parameter
$\tilde p_0$ is not changing.

Here is used mainly a model with spin-half systems, but an analogy of the
parameters \eq{psigm}, Stokes parameters was also considered in optical
models of qubits \cite{JKM}.

\todo{Heisenberg}

\subsection{Two qubits}
\label{sec:twoq}

Let us apply {\em the composition theorem} for system of two qubits. Here we
have {\em complete set} of $36=6^2$ vectors $\ket{\Oi_1^{\mu_1}\Oi_2^{\mu_2}}$
with $\mu_1,\mu_2 = 1,2,3$.
The set is {\em almost perfect}, it can be considered as disjoint
union of nine bases with four elements: $\{\ket{0^{\mu_1}0^{\mu_2}},
\ket{0^{\mu_1}1^{\mu_2}},\ket{1^{\mu_1}0^{\mu_2}},\ket{1^{\mu_1} 1^{\mu_2}}\}$
for nine pairs $(\mu_1,\mu_2)$. In example with spin systems, it corresponds
to nine possible combinations of measurements of two spins
$(S_{\mu_1},S_{\mu_2})$ with four possible combinations of results,
$(\pm\half,\pm\half)$, i.e, $(\half\lambda_{\Oi_1},\half\lambda_{\Oi_2})$.

The set is not {\em perfect}, for example for vector $\ket{0^z 0^z}$
the choice of basis $\{\ket{0^z 0^z},\ket{0^z 1^z},\ket{1^z 0^z},
\ket{1^z 1^z}\}$ is not unique, it is simple to check that
vectors $\{\ket{0^z 0^z},\ket{0^z 1^z},\ket{1^z 0^x}, \ket{1^z 1^x}\}$
are also orthogonal. The example shows that tensor product of two
{\em perfect} sets may be {\em almost perfect}, but not perfect.

It is possible to consider $36$ (linearly dependent) probabilities:
\begin{equation}
 p_{\Oi_1}^{\mu_1}{}_{\Oi_2}^{\mu_2} =
\Tr(\rho \, \P_{\Oi_1}^{\mu_1} \ox \P_{\Oi_2}^{\mu_2}),
\quad \mu_1,\mu_2 = 1,2,3, \quad \Oi_1,\Oi_2 = 0,1
\label{LRprob}
\end{equation}
described by $16$ parameters:
\begin{equation}
\tilde p_{\nu_1\nu_2} = \Tr(\rho\,\sigma^{\nu_1}\ox\sigma^{\nu_2}),
\quad \nu_1,\nu_2 = 0,1,2,3.
\label{LRsigm}
\end{equation}

The Hermitian basis with 16 elements $\frac14\sigma^{\nu_1}\ox\sigma^{\nu_2}$
is \trab-orthonormal (see \Sec{manyq} below) and it is possible to write
\begin{equation}
 \rho = \tfrac{1}{4}\sum_{\nu_1,\nu_2=0}^3{%
 \tilde p_{\nu_1\nu_2}\sigma^{\nu_1}\ox\sigma^{\nu_2}}
\end{equation}

The expressions of $p$ via $\tilde p$ can be simply calculated
\begin{eqnarray}
 p_{\Oi_1}^{\mu_1}{}_{\Oi_2}^{\mu_2} &=&
\Tr(\rho \,\P_{\Oi_1}^{\mu_1} \ox \P_{\Oi_2}^{\mu_2}) =
\tfrac{1}{4}\Tr\BIG({\rho \: (\sigma^0+\lambda_{\Oi_1}\sigma^{\mu_1})\ox
(\sigma^0+\lambda_{\Oi_2}\sigma^{\mu_2})}) \nonumber\\
&=& \tfrac{1}{4}(\tilde p_{00} + \lambda_{\Oi_1}\tilde p_{\mu_1 0}
+ \lambda_{\Oi_2}\tilde p_{0\mu_2} +
\lambda_{\Oi_1}\lambda_{\Oi_2}\tilde p_{\mu_1\mu_2}).
\end{eqnarray}
where $\tilde p_{00} = \Tr \rho = 1$.

On the other hand, to express $\tilde p$ via $p$ it is necessary to
use few kind of formulas (here $\mu_1,\mu_2 \ne 0$):
\begin{eqnarray}
\tilde p_{\mu_1\mu_2} =
\Tr\BIG({\rho\:(\P_0^{\mu_1}\!-\P_1^{\mu_1})\ox(\P_0^{\mu_2}\!-\P_1^{\mu_2})})
= p_0^{\mu_1}{}_0^{\mu_2}\! - p_0^{\mu_1}{}_1^{\mu_2}\!
- p_1^{\mu_1}{}_0^{\mu_2}\! + p_1^{\mu_1}{}_1^{\mu_2}\!\!,&\nonumber\\
\tilde p_{0\mu_2} =
\Tr\BIG({\rho\:(\P_0^{\mu_1}\!+\P_1^{\mu_1})\ox(\P_0^{\mu_2}\!-\P_1^{\mu_2})})
= p_0^{\mu_1}{}_0^{\mu_2}\! - p_0^{\mu_1}{}_1^{\mu_2}\!
+ p_1^{\mu_1}{}_0^{\mu_2}\! - p_1^{\mu_1}{}_1^{\mu_2}\!\!,&\nonumber\\
\tilde p_{\mu_1 0} =
\Tr\BIG({\rho\:(\P_0^{\mu_1}\!-\P_1^{\mu_1})\ox(\P_0^{\mu_2}\!+\P_1^{\mu_2})})
= p_0^{\mu_1}{}_0^{\mu_2}\! + p_0^{\mu_1}{}_1^{\mu_2}\!
- p_1^{\mu_1}{}_0^{\mu_2}\! - p_1^{\mu_1}{}_1^{\mu_2}\!\!,&\nonumber\\
1 = \tilde p_{0 0} =
\Tr\BIG({\rho\:(\P_0^{\mu_1}\!+\P_1^{\mu_1})\ox(\P_0^{\mu_2}\!+\P_1^{\mu_2})})
= p_0^{\mu_1}{}_0^{\mu_2}\! + p_0^{\mu_1}{}_1^{\mu_2}\!
+ p_1^{\mu_1}{}_0^{\mu_2}\! + p_1^{\mu_1}{}_1^{\mu_2}\!\!,&
\end{eqnarray}
there only for nine parameters $\tilde p_{\mu_1\mu_2}$ with
$\mu_1,\mu_2 \ne 0$ it is unique decomposition, for six parameters
with one zero index $\tilde p_{\mu_1 0}$ or $\tilde p_{0\mu_2}$ there are
three different expressions and the last equation shows nine different
ways to decompose unit.

\todo{$A_U$}

It is possible to introduce six probabilities for description of
first subsystem,
\begin{equation}
 p_{\Oi}^{\mu\,(1)} \equiv
\Tr(\rho \, \P_{\Oi}^{\mu} \ox \Id),
\quad \mu = 1,2,3, \quad \Oi = 0,1
\label{Lprob}
\end{equation}
and six for second one,
\begin{equation}
 p_{\Oi}^{\mu\,(2)} \equiv
\Tr(\rho \, \Id \ox \P_{\Oi}^{\mu}),
\quad \mu = 1,2,3, \quad \Oi = 0,1.
\label{Rprob}
\end{equation}

The equations correspond to measurement of only first or second spin
respectively for models with two spin-half systems.

Each such probability can be expressed as sum of two probabilities
described by \eq{LRprob} in three different ways (corresponding to
free parameter $\mu_1$ or $\mu_2$):
\begin{equation}
 p_{\Oi}^{\mu\,(1)} =
 p_{\Oi}^{\mu}{}_{0}^{\mu_2} +  p_{\Oi}^{\mu}{}_{1}^{\mu_2}, \quad
 p_{\Oi}^{\mu\,(2)} =
 p_{0}^{\mu_1}{}_{\Oi}^{\mu} +  p_{1}^{\mu_1}{}_{\Oi}^{\mu}, \quad
 \forall \mu_1,\mu_2 = 1,2,3.
\label{LRexpr}
\end{equation}

It is also possible to introduce similar description with parameters
$\tilde p$, but in this case it is only change of notation:
\begin{equation}
\tilde p^{(1)}_{\nu} = \Tr(\rho\,\sigma^{\nu}\ox\Id) = \tilde p_{\nu 0},
\quad \nu = (0,)\,1,2,3,
\label{Lsigm}
\end{equation}
\begin{equation}
\tilde p^{(2)}_{\nu} = \Tr(\rho\,\Id\ox\sigma^{\nu}) = \tilde p_{0 \nu},
\quad \nu = (0,)\,1,2,3.
\label{Rsigm}
\end{equation}
(where $\tilde p^{(j)}_{0} = \tilde p_{00} = \Tr\rho = 1$, $\forall\,j$).
It is also possible to use the parameters \eq{Lsigm} and \eq{Rsigm} to
express probabilities \eq{Lprob} and \eq{Rprob} respectively:
\begin{equation}
 p_{\Oi}^{\mu\,(j)} = \half(1 + (-1)^{\Oi} \tilde p^{(j)}_{\mu}),
\quad j = 1,2, \quad \mu = 1,2,3, \quad \Oi = 0,1.
\label{jprob}
\end{equation}

Let us consider situation {\em without entanglement}, when:
\begin{equation}
 \rho = \rho_1 \ox \rho_2.
\label{prho}
\end{equation}
(cf. $\ket\psi = \ket{\psi_1} \ox \ket{\psi_2}$ for pure states)

Using identities:
\begin{equation}
 (A \ox B)\,(C \ox D) = (A C) \ox (B D)
\end{equation}
together with
\begin{equation}
 \Tr(A \ox B) = \Tr(A) \Tr(B)
\label{Trox}
\end{equation}
we have for any matrices $A_1, A_2$:
\begin{equation}
 \Tr\BIG({(\rho_1 \ox \rho_2)\, (A_1 \ox A_2)}) =
 \Tr\BIG({(\rho_1\,A_1)\ox(\rho_2\,A_2)}) =
 \Tr(\rho_1\,A_1) \Tr(\rho_2\,A_2).
\label{joint}
\end{equation}

So if systems are not entangled, then due to \eq{prho} and \eq{joint}
it is simple to prove expression for ``independent probabilities''.
Due to definitions \eq{LRprob}, \eq{Lprob}, and \eq{Rprob}, it is possible
to write:
\[
p_{\Oi_1}^{\mu_1}{}_{\Oi_2}^{\mu_2} =
\Tr(\rho_1 \ox \rho_2 \, \P_{\Oi_1}^{\mu_1} \ox \P_{\Oi_2}^{\mu_2}) =
\Tr(\rho_1 \, \P_{\Oi_1}^{\mu_1}) \Tr(\rho_2 \, \P_{\Oi_2}^{\mu_2}),
\]
\[
 p_{\Oi_1}^{\mu_1\,(1)} =
\Tr(\rho_1 \ox \rho_2 \, \P_{\Oi_1}^{\mu_1} \ox \Id) =
\Tr(\rho_1 \, \P_{\Oi_1}^{\mu_1})\Tr(\rho_2) =
\Tr(\rho_1 \, \P_{\Oi_1}^{\mu_1}),
\]
\[
 p_{\Oi_2}^{\mu_2\,(2)} =
\Tr(\rho_1 \ox \rho_2 \, \Id \ox \P_{\Oi_2}^{\mu_2}) =
\Tr(\rho_1)\Tr(\rho_2 \, \P_{\Oi_2}^{\mu_2}) =
\Tr(\rho_2 \, \P_{\Oi_2}^{\mu_2}),
\]
and, finally,
\begin{equation}
p_{\Oi_1}^{\mu_1}{}_{\Oi_2}^{\mu_2} =
p_{\Oi_1}^{\mu_1\,(1)} p_{\Oi_2}^{\mu_2\,(2)}
\label{jntprob}
\end{equation}
Similar expression
\begin{equation}
\tilde p_{\nu_1\nu_2} = \tilde p^{(1)}_{\nu_1} \tilde p^{(2)}_{\nu_2}
\label{jntpar}
\end{equation}
can be proved for definitions \eq{LRsigm}, \eq{Lsigm}, and \eq{Rsigm}.

\subsection{Many qubits}
\label{sec:manyq}

For composite system with
$m$ qubits density matrix is $\rho \in H_{\rm ph}(2^m)$ and it
is possible to use $4^m$ different tensor products of Pauli matrices
as a basis:
\begin{equation}
\mbf{H} = \sum_{\nu_1,\ldots,\nu_m=0}^3
{a_{\nu_1\ldots\nu_m}\sigma^{\nu_1}\ox\cdots\ox\sigma^{\nu_m}},
\quad \mbf{H} \in H(2^m),\quad a_{\nu_1\ldots\nu_m} \in \R.
\label{H2nbas}
\end{equation}

Let us denote
$\sigma^{\nu_1\ldots\nu_m} \equiv \sigma^{\nu_1}\ox\cdots\ox\sigma^{\nu_m}$.
Due to \eq{Trox} it is possible to write:
\begin{equation}
\Tr(\sigma^{\nu_1\ldots\nu_m}\,\sigma^{\mu_1\ldots\mu_m})
= 2^m \delta_{\nu_1 \mu_1} \cdots \delta_{\nu_n \mu_m},
\label{Trsm}
\end{equation}
i.e., the Hermitian basis $2^{-m}\sigma^{\nu_1\ldots\nu_m}$ is
\trab-orthonormal.
An expression for coefficients $a_{\nu_1\ldots\nu_m}$ directly follows
from the \eq{Trsm}:
\begin{equation}
a_{\nu_1\ldots\nu_m}=2^{-m}\Tr(\mbf{H}\,\sigma^{\nu_1\ldots\nu_m}).
\end{equation}
So for density matrix $a_{00\ldots0} = 2^{-m}\Tr\rho = 2^{-m}$.

Using composition of $m$ perfect qubit bases with six vectors discussed
above we can produce basis with $6^m$ components. It is almost perfect
because can be considered as disjoint union of $3^m$ bases with
$2^m$ elements: $\ket{\Oi_1^{\mu_1}\cdots\Oi_m^{\mu_m}}$
marked by $3^m$ different sets $(\mu_1,\ldots,\mu_m)$. It is
similar with description of case $m=2$ above and the counterexample
provided there is enough to show that the set is not perfect for any $m>1$.

Let us denote
$\P_{\Oi_1}^{\mu_1}{}^{\ldots}_{\ldots}{}_{\Oi_m}^{\mu_m} \equiv
\P_{\Oi_1}^{\mu_1}\ox\cdots\ox\P_{\Oi_m}^{\mu_m}$. When $6^m$
probabilities introduced by complete (and almost perfect) set
$\Sv_{6^m}(\Hil_2^{\ox m})$ can be represented as:
\begin{equation}
p_{\Oi_1}^{\mu_1}{}^{\ldots}_{\ldots}{}_{\Oi_m}^{\mu_m} =
\Tr(\rho\,\P_{\Oi_1}^{\mu_1}{}^{\ldots}_{\ldots}{}_{\Oi_m}^{\mu_m}),
\quad \mu_k = 1,2,3, \quad \Oi_k = 0,1, \quad k=1,\dots,m,
\label{mprob}
\end{equation}
described by $4^m$ parameters:
\begin{equation}
\tilde p_{\nu_1\ldots\nu_m} =
\Tr(\rho\,\sigma^{\nu_1\ldots\nu_m}),
\quad \nu_k = 0,1,2,3 , \quad k=1,\dots,m,
\label{msigm}
\end{equation}
where $\tilde p_{0\ldots0} = \Tr\rho = 1$.

In example with spin systems,
$p_{\Oi_1}^{\mu_1}{}^{\ldots}_{\ldots}{}_{\Oi_m}^{\mu_m}$ correspond
to $3^m$ possible combinations of measurements of $m$ spins
$(S_{\mu_1},\ldots,S_{\mu_m})$ with $2^m$ possible combinations of results
$(\half\lambda_{\Oi_1},\ldots,\half\lambda_{\Oi_m})$.

It is possible to introduce six probabilities for each subsystem, similarly
with \eq{Lprob} and \eq{Rprob},
\begin{equation}
 p_{\Oi}^{\mu\,(k)} \equiv
\Tr(\rho \, \Id^{\ox k-1} \ox \P_{\Oi}^{\mu} \ox \Id^{\ox m-k}),
\quad \mu = 1,2,3, \quad \Oi = 0,1, \quad k = 1,\ldots,m
\label{Kprob}
\end{equation}
described by three parameters similarly with \eq{Lsigm} and \eq{Rsigm},
\begin{equation}
\tilde p^{(k)}_{\nu} =
\Tr(\rho\,\Id^{\ox k-1}\ox\sigma^{\nu} \ox \Id^{\ox m-k}) =
\tilde p_{\underbrace{\scriptstyle 0\ldots0}_{k-1}\nu%
\underbrace{\scriptstyle 0\ldots0}_{m-k}}, \quad k = 1,\ldots,m,
\label{ksigm}
\end{equation}
(where $\tilde p^{(k)}_{0} = \tilde p_{0\ldots0} = \Tr\rho = 1$,
$\forall\,j$).
The \eq{jprob} again can be used for calculation of probabilities \eq{Kprob},
\begin{equation}
 p_{\Oi}^{\mu\,(k)} = \half(1 + (-1)^{\Oi} \tilde p^{(k)}_{\mu}),
\quad k = 1,\ldots,m \quad \mu = 1,2,3, \quad \Oi = 0,1,
\label{kprob}
\end{equation}
but analogue of \eq{LRexpr} is too difficult, because it would contain
$2^{m-1}$ terms and may be expressed in $3^{m-1}$ different ways. It is
yet another example of usefulness of parameters $\tilde p$.

Here is also can be considered systems without entanglement,
\begin{equation}
 \rho = \rho_1 \ox \cdots \ox \rho_m = \bigotimes_{k=1}^m \rho_k.
\label{prhom}
\end{equation}

For such states is also can be used analogue of \eq{jntprob}
\begin{equation}
p_{\Oi_1}^{\mu_1}{}^{\ldots}_{\ldots}{}_{\Oi_m}^{\mu_m} =
\prod_{k=1}^m  p_{\Oi_k}^{\mu_k\,(k)}
\quad \mu_k = 1,2,3, \quad \Oi_k = 0,1, \quad k=1,\dots,m,
\label{mjprob}
\end{equation}
and \eq{jntpar}
\begin{equation}
\tilde p_{\nu_1\ldots\nu_m} = \prod_{k=1}^m\tilde p^{(k)}_{\nu_k} ,
\quad \nu_k = 0,1,2,3 , \quad k=1,\dots,m.
\label{mjpar}
\end{equation}

So, systems without entanglement can be described by $6m$ probabilities
\eq{Kprob} instead of $6^m$ \eq{mprob} or $3m$ parameters \eq{ksigm}
instead of $4^m-1$ parameters \eq{msigm}. For systems with entanglement
such simple idea does not work and $6^m$ probabilities of \eq{mprob}
and $4^m$ parameters \eq{msigm} can be considered as some tensors,
\(
\mbf{p_{\Oi}^{\mu}} \equiv
p_{\Oi_1}^{\mu_1}{}^{\ldots}_{\ldots}{}_{\Oi_m}^{\mu_m}
\)
and
\(
\mbf{\tilde p_{\nu}} \equiv \tilde p_{\nu_1\ldots\nu_m}.
\)

Transformations of the tensors due to action of local quantum gates
on initial state is also local. If some quantum gate acts only
on a few ($l=1,2,\ldots$; $l \ll m$) qubits and can be described as
$2^l \times 2^l$ complex matrix, then transformation of tensor
$\mbf{\tilde p_{\nu}}$ can be described by $4^l \times 4^l$
real matrix (and $6^l \times 6^l$ matrix for $\mbf{p_{\Oi}^{\mu}}$).

It is clear after rewriting of expressions with trace like \eq{UrUH}
already used above, i.e,
\begin{equation}
\tilde p'_{\nu_1\ldots\nu_m} = \Tr(U \rho\,U^\hc \sigma^{\nu_1,\ldots,\nu_m})
= \Tr(\rho \,U^\hc \sigma^{\nu_1\ldots\nu_m} U)
\end{equation}
and so if $U$ acts only on few indexes between $\nu_1,\ldots,\nu_m$, then
only the indexes present in matrix $\tilde A_U$ of the transformation
$\tilde \ve p' = \tilde A_U \tilde \ve p$ or, formally,
\begin{equation}
\tilde p'_{\nu_1\ldots\nu_m} = \sum_{\mu_1,\ldots,\mu_m=0}^3{%
\tilde A_{(\nu_1\ldots\nu_m)}^{(\mu_1\ldots\mu_m)}\tilde p_{\mu_1\ldots\mu_m}}
\label{Amany}
\end{equation}
Really it follows directly form expressions for coefficients \eq{tAKJ} that
can be rewritten as
\begin{equation}
\tilde A_{(\nu_1\ldots\nu_m)}^{(\mu_1\ldots\mu_m)} =
 \Tr(\sigma^{\nu_1\ldots\nu_m} U^\hc \sigma^{\mu_1\ldots\mu_m} U).
\end{equation}
For example for two-qubit gate acting on some indexes $j$ and $k$ instead of
\eq{Amany} for $\tilde A_U$ with $4^m$ components it is enough to use only
$4^2=16$ components and write:
\begin{equation}
\tilde p'_{\nu_1\ldots\nu_j\ldots\nu_k\ldots\nu_m} =
\sum_{\mu_j,\mu_k=0}^3{\tilde A_{(\nu_j\nu_k)}^{(\mu_j\mu_k)}%
\tilde p'_{\nu_1\ldots\mu_j\ldots\mu_k\ldots\nu_m}}.
\end{equation}
The similar approach is true for $A_U$ and $\mbf{p_{\Oi}^{\mu}}$.

Such probabilistic description of quantum circuits is not more complex, than
usual description with pure states, and for description of quantum channels
with mixed states it is even more simple, because here is not necessary to
use some special decomposition for ``superoperators'', they are also may be
expressed by one matrix with same size as for ``usual'' operators.

Here again was rather used model with $m$ spin-half systems, but parameters
similar with $\mbf{\tilde p_\nu}$ was also used as ``multiple beam Stokes
parameters'' in rather optical framework \cite{JKM}.

\todo{Heisenberg}

\section{Conclusion and bibliographical notes}

The set of problems discussed in present paper is part of more general task
of complete description of quantum systems using some set of observables. The
task had a long history and very wide range of different approaches and it is
not possible here to review that, but together with references used in main
body of the paper it should be mentioned some alternative approaches
\cite{CMS,ShorPOVM,DBK} together with references therein.  Present paper
itself appears initially in rather specific circumstances of discussions
related with macroscopic realization of quantum logic \cite{GZ90,GZ92,BG,GR99}
with later development in more general framework related with quantum
computations.

\section*{Acknowledgements}

Author is grateful to A. Grib, V. Bubovich, R. Zapatrin for discussions
and also to many other colleagues for participation in some seminars about
year ago devoted to consideration of initial version\footnote{See also
\footref{hist} on page \pageref{hist}.} of present work in
A.A.Friedmann Laboratory for Theoretical Physics.

\end{document}